\documentclass[12pt,a4paper]{article}
\usepackage[left=1.5cm,top=1cm,right=1.5cm]{geometry}
\usepackage{setspace}
\usepackage[pdftex]{graphicx}           
\usepackage{hyperref}
\usepackage{multirow}
\usepackage{amsmath}
\usepackage{ulem}

\title{Comparing intermittency and network measurements of words and their dependency on authorship}

\author{Diego Raphael Amancio\\
\small Institute of Physics of S\~ao Carlos\\[-0.8ex]
\small University of S\~ao Paulo, P. O. Box 369, Postal Code 13560-970 \\
\small S\~ao Carlos, S\~ao Paulo, Brazil \\
\small \texttt{diego.amancio@usp.br}\\
\and
Eduardo G. Altmann\\
\small Max Planck Institute for the Physics of Complex Systems\\
\small Dresden, Germany\\
\small \texttt{edugalt@pks.mpg.de}\\
\and
Osvaldo Novais Oliveira Jr.\\
\small Institute of Physics of S\~ao Carlos\\[-0.8ex]
\small University of S\~ao Paulo, P. O. Box 369, Postal Code 13560-970 \\
\small S\~ao Carlos, S\~ao Paulo, Brazil \\
\small \texttt{chu@ifsc.usp.br}\\
\and
Luciano da Fontoura Costa\\
\small Institute of Physics of S\~ao Carlos\\[-0.8ex]
\small University of S\~ao Paulo, P. O. Box 369, Postal Code 13560-970 \\
\small S\~ao Carlos, S\~ao Paulo, Brazil \\
\small \texttt{ldfcosta@gmail.com}\
}

\date{{\sc Published as: New Journal of Physics, 123024 (2011)} \\
\url{http://dx.doi.org/10.1088/1367-2630/13/12/123024}\\
\ \ \\
{\sc Supplementary Information at:}\\ 
\url{http://iopscience.iop.org/1367-2630/13/12/123024/media}
}

\begin{document}
\maketitle

\onehalfspacing

\newpage

\tableofcontents

\newpage

\begin{abstract}
 Many features from texts and languages can now be inferred from statistical analyses using concepts from complex networks and dynamical
    systems. In this paper we quantify how topological properties of word co-occurrence networks and intermittency (or burstiness) in word
    distribution depend on the style of authors. Our database contains 40 books from 8 authors who lived in the 19th and 20th centuries, for
    which the following network measurements were obtained: clustering coefficient, average shortest path lengths, and betweenness. We found
    that the two factors with stronger dependency on the authors were the skewness in the distribution of word intermittency and the average
    shortest paths. Other factors such as the betweeness and the Zipf's law exponent show only weak dependency on authorship. Also assessed was the contribution from each measurement to authorship recognition using three machine learning methods. The best performance was a ca. 65~\% accuracy upon combining complex network and intermittency features with the nearest neighbor algorithm. From a detailed analysis of the interdependence of the various metrics it is concluded that the methods used here are complementary for providing short- and long-scale perspectives of texts, which are useful for applications such as identification of topical words and information retrieval.
\end{abstract}

\newpage
\section{Introduction}

The application of ideas from statistical physics to text analysis has a long tradition since Shannon's usage of entropy as the central concept in information theory~\cite{Shannon}. In recent years, physicists have proposed new approaches based on concepts from complex networks~\cite{sole,cancho2,leasteffort,cancho,patt,lantiq1,trad1,trad2,summ,spellcheck,poesias,fatoficcao,iss,masucci}
and dynamical systems~\cite{Montemurro,Berryman,Altmann,Goh,Allegrini}.
In the former, text is represented as complex networks with words (nodes) being connected (links) using procedures depending on their syntactic or semantic relationships~\cite{sole}. Several of these networks share topological properties such as the scaling in the degree~\cite{cancho2,leasteffort} and the small world feature~\cite{cancho,patt}. The
co-occurrence networks, where adjacent words are linked to each other, are probably the most popular for applications owing to their ability
to capture important syntactic and semantic aspects of texts with a straightforward construction procedure. These networks were employed to
evaluate writing quality~\cite{lantiq1} and machine translations~\cite{trad1,trad2}, to generate and evaluate  summaries~\cite{summ}, to
construct spell checkers~\cite{spellcheck}, to recognize patterns in poetry~\cite{poesias} and prose~\cite{fatoficcao,iss} and to study
general properties of written language~\cite{masucci}. While co-occurrence networks focus mainly on short scales, an increasingly popular
approach addresses longer text scales~\cite{Montemurro,Berryman,Altmann,Herrera,Carpena,Katz,Ortuno}. The usefulness of this latter approach
stems from the finding that topical words are unevenly distributed along the text when compared to a random process or to function words. This observation can be quantitatively investigated using different analogies and measures familiar to the communities of statistical physics and dynamical systems, including level statistics~\cite{Carpena,Ortuno}, burstiness~\cite{Berryman,Altmann,Goh},  entropy~\cite{Herrera}, and intermittency measures~\cite{Allegrini}.  The author dependency on the features mentioned above has been noticed~\cite{iss,Berryman}, but little work has been devoted to quantify the extent of this dependency and to test its usefulness to the automatic detection of authors.

In the field of authorship recognition (or stylometry), one tries to identify the author of documents whose identity is lacking~\cite{oakes}.
Some simple quantitative proposals, such as the use of word length to distinguish between authors, go back to the mid 19th century (see
Ref.~\cite{Tankard} for a historical account). One important recent contribution was given by Mosteller and Wallace~\cite{Mosteller} who showed that the frequency of function words (such as ``any'', ``from'', ``an'',
``there'' and ``upon'') can be used to characterize the style of authors. This feature is so strong that even letter pair frequencies can provide a good distinguishability between authors~\cite{Tankard}. Frequent words are also responsible for the success of the approach -- proposed and investigated by physicists -- that consist in quantifying the similarity between two books based on the distance between their word-frequency rankings~\cite{havlin,vilensky,yang}.  More recently, new features have been proposed: word length, sentence length; frequency of punctuation marks and contractions; frequency of graphemes, collocations and words. A summary of these recent results is given in Ref.~\cite{Grieve}.

In this paper we investigate how the metrics of complex networks and intermittency -- familiar to physicists -- depend on the style of authors.
We start quantifying the metrics for each word (Sec.~\ref{sec.2}). These are used in the definition of global features for each book that
are tested according to their efficiency in algorithms of authorship classification (Sec.~\ref{sec.3}). Finally, we discuss the importance
of each feature (Sec.~\ref{sec.4}).
 The primarily goal of this paper is {\bf not} to improve state-of-the-art methods of automatic authorship recognition. Instead, we wish to estimate the dependency on authorsihp of the selected metrics.
We evaluate our results using authorship classification tests because they provide a statistical rigorous method to quantify
the importance of different features. Nevertheless, our study reveals interesting insights which are potentially
useful in real applications and therefore we include a comparison to more traditional statistical natural language methods
(that increase the efficiency from $62.5\%$ to $90.0\%$ correct attributions, see Sec.~\ref{ssec.comparison}).

\section{Statistical quantification of the role of words in texts} \label{matmet}\label{sec.2}

\subsection{Database} \label{ssec.books}

Our database contains 5 novels from each of 8 authors who lived between 1809 and 1975, which are available in an online repository
(http://www.gutenberg.org/). The list of books is summarized in Sec. 1 of the Supplementary Information (SI). To avoid effects from the
length of the texts, each book was limited to their first 18,200 tokens, which corresponds to the length of the shortest book. In the
remainder of this section we illustrate our results using the book ``The Adventures of Sally'', by P. G. Wodehouse. The results for all
books appear in (SI)-Sec.~3 and are discussed in Sec.~\ref{sec.recognition} below.

\subsubsection*{Pre-processing of the text.}

Before extracting complex networks and intermittency measurements from the texts, some preprocessing steps were applied. Initially, a
pre-compiled list of stopwords including articles, prepositions and adverbs were removed from the text (see SI-Sec.~2). Previous work for
recognizing authorship used the frequency of function words, but we decided not to use them in our study because we are interested in the
interrelation between words with a pronounced semantic content. This procedure has been employed in many works (see
e.g. Refs.~\cite{lantiq1,trad1,trad2,poesias,masucci}) and it is crucial to determine how these techniques depend on the writing style of
each author. Next, a lemmatization step was applied to the remaining words using the MXPost part-of-speech Tagger based on the Ratnaparki's
model~\cite{ratna}. Table \ref{tab.1} exemplifies the application of these pre-processing steps. With this
standardization, we grouped together all words referring to a same concept, despite the differences in flexion.

\subsection{Network measurements} \label{modeling}\label{sec.network}

\begin{figure}
		\begin{center}
			\includegraphics[width=1\textwidth]{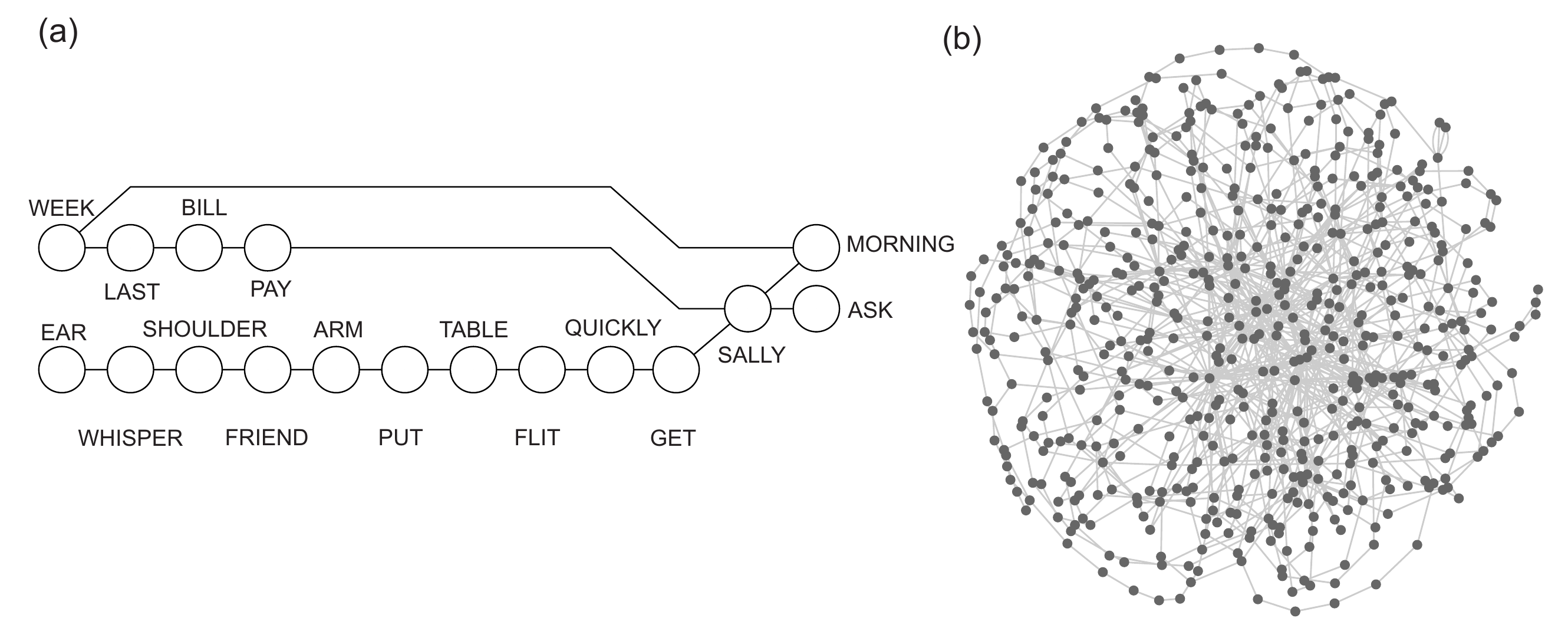}
		\end{center}
		\caption[\it]{\it Example of networks: (a) the subgraph obtained for the sentences shown in Table~\ref{tab.1}; and (b) the global network
                  obtained from the first $1,000$ associations of the same book.}
		\label{fig1}
\end{figure}

Complex networks have been used to characterize different properties of languages~\cite{cancho,patt,lantiq1,trad1,trad2,summ,iss,lang}. Here we are interested in author-specific characteristics and therefore we adopt a network description based on word co-occurrence~\cite{cancho,lantiq1,trad1,trad2,iss,lang}, where nodes are words and links are established between subsequent words. This procedure is illustrated in Fig.~\ref{fig1}. The network is defined by a set $V$ = \{$v_1$, $v_2$, $\ldots$ $v_n$\} of vertices, and a set $E$ of edges and is represented as a nonsymmetric weighted matrix $W$. By construction, $W$ is a square matrix of size $n$, where $n$ is the number of distinct words after the pre-processing step. The elements $w_{ij}$ of $W$ indicate the strength that $v_i$ is connected to $v_j$ ($v_i$ $\rightarrow$ $v_j$), i.e. the number of times word $v_j$ appears immediately after word $v_i$.
Additionally, we used the non-weighted and undirected network corresponding to $W$, denoted by the matrix $A$ whose elements $a_ {ij}$ = 1 if the words represented by the vertices $v_i$ and $v_j$ appeared as neighbors at least once in the text. Otherwise, $a_{ij}$ = 0.

We shall use the statistical properties of complex network measurements in the networks $W$ and $A$. In this section we discuss the word-specific local measurements and in Sec.~\ref{ssec.localtoglobal} we show how to connect them to obtain a global characterization of the network. The number of occurrences $N_i$ of each word $i$ represented by node $v_i$ is
\begin{equation} \label{frequencia}
	N_i = s^{in}_i = s^{out}_i = \sum_{j} w_{ij},
\end{equation}
where $s^{in}$ and $s^{out}$ are respectively the weight (or strength~\cite{strengthRef,luciano}) of the ingoing and outgoing edges of node
$v_i$\footnote{The triple equality in Eq. (\ref{frequencia}) is not valid for the first and for the last word in the text. The correct
  expression for the first word is $N_i =  k_{out}(i) = \sum_{j} w_{ij}$ and for the last word it is $N_i =  k_{in}(i) = \sum_{i}
  w_{ij}$.}. Therefore, the degree of each node~$v_i$ is the frequency of appearance of word~$v_i$ and the degree distribution of~$W$ is
proportional to the normalized frequency ($f_i=N_i/N_T, N_T=\sum_i N_i$) distribution of words, given by the Zipf's
law~\cite{zipf,bernhardsson}. Below we discuss three typical measurements: clustering coefficient, average shortest path length, and betweenness centrality.

\begin{table}
    \centering
	\caption{\label{tab.1} Example of the pre-processing steps applied to the texts. An extract (first column) obtained from the book ``The Adventures of Sally'', by Pelham Grenville Wodehouse is shown after the removal of the stopwords (second column) and after lemmatization (third column).}
		\begin{tabular}{lll}
			\hline
			\textbf{Original}	&\textbf{Without stopwords}	&\textbf{After lemmatization}  \\
			\hline
			What's that ? asked Sally. 		&  asked Sally 					& ask Sally 					\\
			Pay my bill for last week,			&  pay bill last week			& pay bill last week 		\\
			due this morning. Sally got up 	&  morning Sally got  			& morning Sally get  		\\
			quickly, and flitting down the   &  quickly flitting 				& quickly flit 				\\
			table, put her arm round her 		&  table put arm 					& table put arm 				 \\
		   friend's shoulder and whispered  &  friend shoulder whispered	& friend shoulder whisper	\\
			in her ear.								&  ear								& ear								 \\
			
			\hline
		\end{tabular}
\end{table}

\subsubsection*{Clustering Coefficient C.}

The clustering coefficient ($C$) measures the probability that the neighbors of a given vertex $v_i$ are connected. This measurement has been widely employed in complex networks, e.g. to verify the presence of communities~\cite{newman1,btwcom,newmanbook} and to distinguish random networks from other small world networks~\cite{cancho,swn}. Traditionally, the clustering coefficient is defined without considering weights or directions as:
\begin{equation} \label{aglomeracao}
	C_i = 3 \frac{ \sum_{k > j > i} a_{ij} a_{ik} a_{jk} }{ \sum_{k > j > i} a_{ij} a_{ik} + a_{ji} a_{jk} + a_{ki} a_{kj} },
\end{equation}
which is equivalent to the fraction of the number of triangles among all possible triads of connected nodes, and therefore ranges from 0 to 1.
With regard to the interpretation of this measure, Ferrer i Cancho and Sol\'e~\cite{cancho} found that the clustering coefficient of
networks representing text was much larger than the one expected just by chance (i.e., the value expected for the corresponding random
networks).

We singled out the words (and their neighbors) with the highest and lowest values of clustering coefficient in the book {\it ``The Adventures of Sally''}, by P. G. Wodehouse, which are shown in Table  \ref{tab.32} for the frequency $N$ = 5\footnote{The words with $N_i < 5$ were considered to lack statistics and were disregarded in all the analysis involving the clustering~$C$.}. From the definition, one expects that words with highest $C$ to have neighbors also connected to each other.
This is the case of the words ``sand'' and ``excitement''. On the other hand, words whose neighbors are not related to each other at all display low values of $C$ (e.g. there is no link between the neighbors of ``full'' or between the neighbors of ``high''). Qualitatively, the clustering coefficient quantifies how words are connected to specific contexts.
Indeed, the words ``sand'' and ``excitement'' tend to be more restricted to a specific context, while ``full'' tends to appear in a myriad of contexts. Therefore, it seems that the clustering coefficient can be useful to detect authorship by quantifying the tendency of using semantic-specific or generic words.

\begin{table}
\centering
\caption{\label{tab.32} Words of the book ``The Adventures of Sally'' with the highest and lowest clustering coefficients (the average clustering $\langle C \rangle$ = 0.085), for words with $N_i$ = 5. The five words with $C = 0$ were randomly selected among the 18 words with $N = 5$ and $C = 0$.}
\begin{tabular}{@{}llll}
\hline
\textbf{Word} & \textbf{Neighbors} & \textbf{N} &\textbf{C} \\
\hline
shortly	& twelve, see, say, Sally, news,						&	5	&	0.27	\\
			& never, heaven, find, enter, Carmylle				&		&	\\
excitement	& thing, suppressed, Sally, mince, last, 						&	5	&	0.25	\\
			& can, come, bristle, brief, apart				&		&	\\
sand	   & watch, want, sit, shuffling, seat,  &	5	&	0.18	\\
			& 	here, golden, first, dark, Roville\\
nose		&	voice, tip, sort, smut, smooth	&	5	&	0.18	\\
			&	Sally, oh, glance, tell, come		&		&		\\
country	&	time, still, somewhere, say, place	&	5	&	0.18	\\	
			&	may, happen, great, glorious	\\
startle	& shy, seem, mill, little, gratify, 						&	5	&	0.00 \\
			& gather, first, everyday, displeased, considerably	& 		&		  \\
high		& recess, mouth, motive, lapse, figure,					&	5	&	0.00	\\
			& even, disposal, critical, collar, check				&		&		\\
gold		& voice, spin, pencil, loan, knob, 						&	5	&	0.00	\\
			& information, high, heavy, frame, buy				&		&		\\
gift		& tongue, take, sort, potential, mean,				&	5	&	0.00	\\
			& few, easily, compensating, blessing, acquire	&		&		\\
full		& Tuesday, peal, later, home, happy,					&	5	&	0.00	\\
			& gratitude, gleaming, glance, color, battle			&		&	\\
\hline
\end{tabular}
\end{table}

\subsubsection*{Average Shortest Path Length~$L$.}

		A shortest path (or geodesic path) between two nodes is defined as the path whose sum of edge weights is minimum. We start defining $d_{ij}$ as the length of the shortest path between $v_i$ and $v_j$ (in this case $A$ is employed). Then the average shortest (or geodesic) path length for $v_i$ ($L_i$) is the average shortest path to all other ($n$-1) nodes of the network:
\begin{equation} \label{geodesico}
	L_i = \frac{1}{n-1} \sum_{j=1}^{n} d_{ij},
\end{equation}
which takes low values if $v_i$ is close to the other nodes.

The words with the lowest $L$ include the characters ``Sally'' ($L = 2.35, N = 347$) and ``Fillmore'' ($L = 2.51, N = 138$), in addition to
high-frequency words, such as ``say'' ($L = 2.45, N = 349$), ``good'' ($L = 2.46, N = 107$) and ``man'' ($L = 2.50, N = 193$). As for the
words with the highest $L$, we found: ``white-clad'' ($L$ =  6.33, $N = 1$), ``affability'' ($L = 6.31, N = 1$), ``whirl'' ($L =  5.89, N =
1$), ``jazz'' ($L = 5.87, N = 1$), ``war-aims'' ($L = 5.84, N = 1$). Interestingly, all these 5 words appeared only once in the text,
indicating that one of the reasons for a high $L$ could be the low frequency $N$. However, $L$ is not only a consequence of the
frequency~$N$ of the words, as low frequency words can also take low values of $L$. This is illustrated in Table \ref{tab.splength}, which
compares words with the same $N$ but different $L$. The frequency has a limited influence on $L$, with a Pearson correlation Corr($L$,$N$) =
-0.36 calculated over all words. Actually, the determining factor is the neighborhood of the word. To understand why this happens, consider
the words ``affability'' and ``repose''. While the former has as neighbors the words ``jaunty'' ($N = 1$) and ``white-clad'' ($N = 1$), the
latter has as neighbors the words ``Sally'' ($N = 347$) and night ($N = 20$). Therefore, one may infer that $L$ actually quantifies the
importance of a word according to its distance to the most frequent words. Since we removed stopwords, the shortest path may be thought of
as quantifying the distance from a word to the core-content words of the book.

\begin{table}
\centering
\caption{\label{tab.splength} Comparing the average shortest path length $L$ for words with the same frequency $N$ of the book ``The Adventures of Sally''. For a given $N$, $L$ may vary widely, which shows the dependency of $L$ on the neighborhood connectivity.}
\begin{tabular}{@{}lll|lll}
\hline
\textbf{Word} & \textbf{$N_i$} & \textbf{$L_i$} & \textbf{Word} & \textbf{$N_i$} & \textbf{$L_i$} \\
\hline
{\it red} & $5$ & $3.71$&{\it earth} & $5$ & $2.99$ \\
{\it shudder} & $4$ & $3.97$&{\it lucky} & $4$ & $3.00$ \\
{\it Maxwell} &  $3$ & $5.55$ & {\it funny} & $3$ & $3.10$\\
{\it dark} & $2$ & $5.15$ & {\it kiss} & $2$ & $3.08$\\
{\it affability} & $1$ & $6.34$ & {\it repose} & $1$ & $3.11$ \\
\hline
\end{tabular}
\end{table}

\subsubsection*{Betweenness.} Betweenness $B$ is a measurement of centrality, with higher values being assigned to the nodes considered as the most relevant in terms of linking different words.
In other words, with $B$ one attempts to quantify the frequency of access of each node, assuming that a given target node in the network is reached from a specific source node via shortest paths. Betweenness is defined as follows. Let $\eta_{st}^i$ be the number of distinct shortest paths between the source node $v_s$ and the target node $v_t$ that pass through the node $v_i$. If $g_{st}$ is the total number of shortest paths between $v_s$ and $v_t$, then $B_i$ is given by:
\begin{equation}\label{eq.bi}
	B_i = \sum_{s} \sum_{t} \frac{\eta_{st}^i}{g_{st}}.
\end{equation}

In the context of text analysis, high frequency words tend to have high $B$. However, some words may play the role of articulation points by
linking concepts related to distinct communities. To illustrate this, we show in Table \ref{tab.btw} that words with similar $N$ may take
very different $B$. A comparison between the left and right columns suggests that words with high $B$ connect concepts because of their probable appearance in various contexts. Therefore, similarly to the clustering coefficient $C$, the betweenness centrality $B$ seems to quantify the variety of contexts in which a word can appear. Note, however, that $B$ is based on a global connectivity pattern, in contrast to $C$.

\begin{table}
\centering
\caption{\label{tab.btw} Comparing the betweenness $B$ for words of the book ``The Adventures of Sally'' with the same frequency $N$. For a given $N$, the betweenness may vary widely, since low frequency words may have high betweenness as they may appear in different contexts.}
\begin{tabular}{@{}lll|lll}
\hline
\textbf{Word} & \textbf{$N_i$} & \textbf{$B_i$} & \textbf{Word} & \textbf{$N_i$} & \textbf{$B_i$} \\
\hline
{\it say}  & $349$ & $745,634$ & {\it Sally} 	& $347$ & $1,192,881$ 	\\
{\it know} & $143$ & $243,357$ & {\it Fillmore} & $138$ & $393,955$ 		\\
{\it tell} & $65$  & $53,904$  & {\it Gerald} 	& $62$  & $108,528$ 		\\
{\it allow} & $20$ & $15,816$  & {\it Roville}  & $21$  & $32,449$  		\\
{\it heaven} & $10$ & $1,147$  & {\it second} 	& $10$  & $22,004$ 		\\
{\it rugger} & $5$ & $855$   	 & {\it worthy}   & $5$   & $10,503$ 		\\
{\it fish} & $4$ & $174$ 		 & {\it spectator} & $4$ & $14,746 $\\
{\it paper-knife} & $3$ & $233$ &{\it group} & $3$ & $8,320$ \\
{\it worship} & $2$ & $44$ & {\it sell} & $2$ & $8,346$ \\
{\it thaw} & $1$ & $11$ & {\it price} & $1$ & $8,295$ \\
\hline
\end{tabular}
\end{table}

\subsection{Intermittency measurements}\label{sec.intermittency}

The uneven distribution of words across {\bf different} documents is an essential feature exploited in Statistical Natural Language Processing. For instance, by investigating words appearing over concentrated in specific documents (when compared to their overall frequency) one can detect keywords, topics, and authorship~\cite{Mosteller,Manning}. This is the basic idea of the {\it term frequency - inverse document   frequency} (TF-IDF) and related measures that are also at the core of search engines~\cite{Manning}. However, there are numerous
situations where the comparison to a general database is not available or is not interesting. For instance, when authorship has to be attributed without previous knowledge of texts written by the potential authors. Here we approach these problems by taking advantage of the finding that words are unevenly distributed not only {\bf across} documents but also {\bf within}
them~\cite{Montemurro,Berryman,Altmann,Herrera,Carpena,Katz,Ortuno}.

The quantification of the uneven distribution of words has been proposed based on measures commonly used by
physicists~\cite{Montemurro,Ortuno}. Following Refs.~\cite{Herrera,Carpena,Ortuno}, we use the statistics of recurrence times, a standard
quantification of intermittency or burstiness in time series~\cite{Altmann,Goh}. In texts, time is counted by the number of words and for
each word~$i$ the recurrence time~$T_j$ is defined as the number of words between two successive occurrences of $i$ (the $j$
and~$j+1$ occurrence) plus one.
For instance, the  recurrence times for the word {\it ``the''} in the previous sentence are~$T_1=9$ and $T_2=7$.
A word that appears ~$N_i$ times in a text of size~$N_T$ leads to a sequence of $N_T-1$ inter-occurrence times~$\{T_1,T_2,...,T_{N_T-1}\}$. In order to incorporate also the time until the
first~$T_f$ and after the last~$T_l$ occurrence of the word, we consider $T_N=T_f+T_l$.
In this case~$\overline{T} = N_T/N_i$, where the overline denotes
average over the different~$T_j$'s. Note that the mean recurrence time~$\overline{T_i}$ gives no additional information than the
frequency~$N_i$. The intermittency of the word appears in the variance
of~$T_j$'s around~$\overline{T}$ and can be quantified by~$I\equiv \sigma_T/\overline{T}$ where~$\sigma_T=\sqrt{\overline{T^2}  - \overline{T}^2}$.  Randomly distributed words in the text have~$I =1$ (in the limit
of large~$N$ and small~$N_i/N_T$), intermittent words have~$I > 1$, and
words appearing in regular intervals have $I <1$. We calculate the intermittency measure~$I_i=\sigma_T/\overline{T}$ for all words with
$N_i\geq 5$ in each of the books (filtered texts) described in Sec.~\ref{ssec.books}. The words with $N_i < 5$ were considered to lack
statistics and were disregarded.

\begin{table}
\centering
\caption{\label{tab.intermit}
In the book  {\it ``The Adventures of Sally''}, by P. G. Wodehouse, there are ~$N_T=15,173$ words (tokens), $3,657$ different word types, and $716$ words with~$N_i\geq 5$. The $5$ words with highest $\sigma_T/\overline{T}$ are shown in the left part of the table.  For comparison, in the right we show for each of these words another word with the closest frequency.}
\begin{tabular}{@{}lll|lll}
\hline
\textbf{Word} & \textbf{$N_i$} & \textbf{$I_i\equiv\sigma_T/\overline{T}$} & \textbf{Word} & \textbf{$N_i$} & \textbf{$I\equiv\sigma_T/\overline{T}$}\\
\hline
{\it jules} & $26$ & $4.31$&{\it turn} & $26$ & $1.55$ \\
{\it hobson} & $31$ & $4.09$&{\it here} & $31$ & $1.35$ \\
{\it ginger} &  $115$ & $3.86$&{\it get} & $117$ & $1.24$\\
{\it carmyle} & $54$ & $3.60$&{\it feel} & $53$ & $0.87$\\
{\it bunbury} & $20$ & $3.59$&{\it people} & $20$ & $1.39$ \\
\hline
\end{tabular}
\end{table}

In Table \ref{tab.intermit} we compare words with highest $I = \sigma/\overline{T}$ to words with similar frequency. It is clear that the most intermittent words (largest $\sigma_T/\overline{T}$) are topical
words (e.g., name of characters and locations), regardless of their frequency. Indeed,~$15$ out of the~$16$ most intermittent words.
are directly connected to specific characters. A similar behavior is observed in all books of our database. Intermittency is therefore a good characterization of topical words that in turn plays an important role in the author-specific characteristic of the texts.
The relationship between $\sigma_T/\overline{T}$ and the function of the words has been investigated in detail in Refs.~\cite{Montemurro,Berryman,Altmann,Herrera,Carpena,Ortuno}. In the next section we explore the fact that these properties are also author specific~\cite{Berryman}.

\section{Evaluating the author dependency}\label{sec.recognition}\label{sec.3}

\subsection{From properties of words to properties of books}\label{ssec.localtoglobal}

\begin{figure}
		\begin{center}
			\includegraphics[width=1\textwidth]{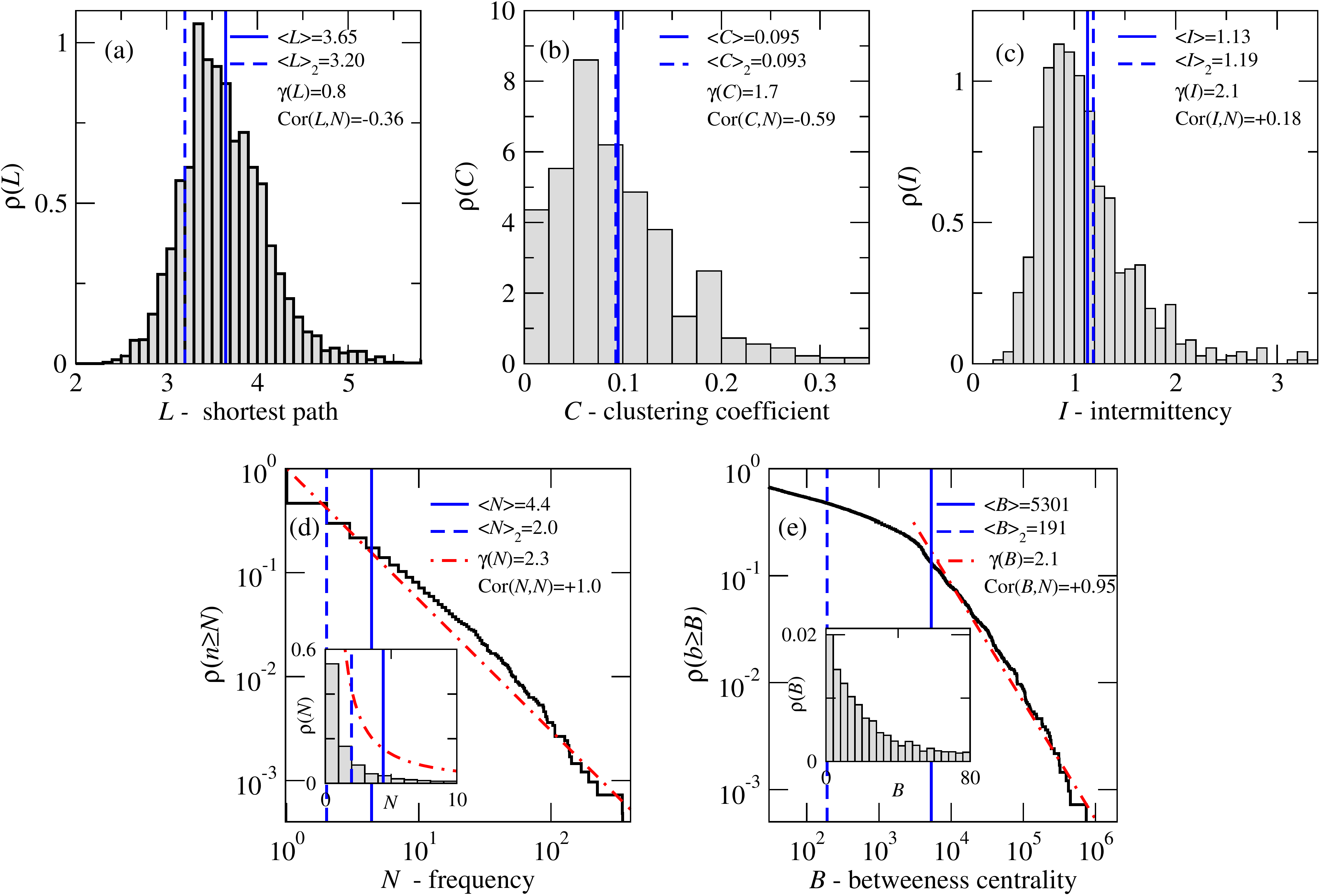}
		\end{center}
		\caption[\it]{\it Probability density function~$\rho(X)$ obtained from the different words of the book {\it ``The Adventures of Sally''}, by P. G. Wodehouse. (a) $X=L$ shortest path, (b) $X=C$ clustering coefficient, (c) $X=I$ intermittency, (d) $X=N$ frequency, and (e) $X=B$ betweenness. In (d) and (e) the cumulative distribution~$\rho(x \ge X)\equiv \int_X^\infty \rho(x) dx$ is shown, with the density~$\rho(X)$ depicted in the inset. The legends indicate the features defined in Eqs.~(\ref{eq.average})-(\ref{eq.tail}) obtained for these distributions, and Cor$(X,N)$ indicates the Pearson correlation coefficient between $X$ and $N$ calculated over all words.}
		\label{fig4}
\end{figure}

 In the previous section we introduced five quantities characterizing  properties of words in the text: frequency ($N$),   average shortest path length ($L$), betweenness ($B$), clustering coefficient ($C$), and intermittency ($I=\{\sigma_T/\overline{T}\}$). The values of these  quantities for all words in the  books in our database can be found in SI-Sec.~3. We now analyze the global distribution of these measurements for all the words in a given book by plotting the empirical probability density function~$\rho(X)$ for the measurements $X = \{N, L, B, C, I\}$.  Fig.~\ref{fig4} shows the results for one book, and similar distributions were obtained for the other books. The shortest path~$L$, clustering~$C$, and intermittency~$I$ have a well defined peak and width (akin to a Gaussian distribution), but the frequency~$N$ and betweenness~$B$ have broad tail distributions (as in power law distributions $\rho(X) \sim X^{-\alpha}$). The tail in~$N$ corresponds to the well-known Zipf's law, which also appears in~$B$ as expected from the large correlation between $B$ and $N$ (Corr($B$,$N$) = $0.95$ in the book of Fig.~\ref{fig4}). With the two different behaviors we propose two sets of measurements, one for~$X=\{L,C,I\}$ and another for~$X=\{N,B\}$.

Our goal is to obtain quantities characterizing important features of these distributions to be used as global measurements of the
books. The most natural choice is the average value~$\langle X \rangle$, where $\langle \ldots \rangle \equiv \frac{1}{M} \sum_{i=1}^M
\ldots$ corresponds to an average over the M different words. For the network measures~$L,C,I$ this corresponds to the average values over
nodes, a quantity considered as characteristic of the network~\cite{lantiq1,poesias,iss,survey}. For $X=\{N,B\}$, the highly frequent words
contribute strongly to~$\langle X \rangle$ due to the long tails. To compensate for this effect, we consider also a modified average defined
as~$\langle X \rangle_2 \equiv\langle \log X \rangle$ for $X=\{N,B\}$. For ~$X=\{L,C,I\}$ the opposite is true, i.e., $\langle X \rangle$ is
dominated by the large number of low frequency words. Accordingly, we introduce a modified average as $\langle X \rangle_2 \propto \sum_i
X_i \log N_i$, i.e., a weighted average with weights proportional to the logarithm of the frequency. The quantities $\langle X \rangle$ and
$\langle X\rangle_2$ are expected to give a good account of ``typical'' values of~$X$. However, in Sec.~\ref{sec.2} we mentioned that
important information is conveyed by words with large ~$X$, i.e., in the tails of the distributions shown in Fig.~\ref{fig4}. In order to
characterize the fat-tail distributions of~$X=\{N,B\}$, we used the coefficient $\alpha_X$ of a power-law fit to the tails
of~$\rho(X)$\footnote{The fitting was performed to the cumulative distribution with logarithmic binning size, as suggested in
  Ref.~\cite{Bauke}. A cut-off $X>3\; 10^3$ was used for $X=B$ (see Fig.~\ref{fig4}e), no cut-off was used for~$X=N$.}. An additional
motivation for using $\alpha_N$ comes from the suggestion in Ref.~\cite{facto} that it
serves as a quantification of the style of texts. The large values of ~$X=\{L,C,I\}$ were characterized by calculating the skewness of~$\rho(X)$, a measure of the asymmetry of the distribution. In  summary, the three features we use for each of the five quantities~$X = \{N, B, L, C, I\}$ are:
\begin{equation}\label{eq.average}
\text{Average value: } \;\; \langle X \rangle \;\; \text{ for } X=\{N,B,L,C,I\}.
\end{equation}
\begin{equation}\label{eq.mavg}
\text{Modified average: } \;\; \langle X \rangle_2 = \left\{ \begin{array}{ll} \langle \log(X)\rangle  & \text{ for } X=\{N,B\}, \\
 \langle X \log N \rangle/\langle \log N \rangle  & \text{ for } X=\{L,C,I\}. \\
\end{array}
\right.
\end{equation}
\begin{equation}\label{eq.tail}
\text{Right tail: } \;\; \gamma(X) = \left\{ \begin{array}{ll} \alpha \text{ in } X^{-\alpha} & \text{ for } X=\{N,B\}, \\
 \text{ skewness}(X)\equiv \langle \left( \frac{X-\langle X \rangle}{\sigma_X} \right)^3 \rangle & \text{ for } X=\{L,C,I\}. \\
\end{array}
\right.
\end{equation}
These features are given in Fig.~\ref{fig4} for one book (see SI-Sec.~3 for all~$40$ books). Obviously, the choice of the quantities above
is inevitably arbitrary. Our choice was intended to capture features of the distribution, rather than giving a parametric description of the
full distribution. In particular, the power-law fit in Eq.~(\ref{eq.tail}) does not intend to fully describe the distributions, as
  apparent in Fig.~\ref{fig4}(d,e).

\subsection{Machine learning methods and evaluation} \label{mlalg}

In order to quantify the ability of the features described above to distinguish between authors, we employ machine learning algorithms which induce classifiers from a training database. The robustness of our results is tested with three widely used algorithms based on different principles. The first is known as C4.5~\cite{quinlan}, and generates decision trees based on the information gained by each feature; the second algorithm is the
Naive Bayes~\cite{bayes}, which is based on the Bayes theorem; and the third and simplest algorithm is the Nearest Neighbor~\cite{knn},
which classifies an unknown instance according to the nearest neighbor of that instance in a normalized space involving all features.
For more details, see SI-Sec.~4.

\subsection{Efficiency of the classification}

We consider the problem of distinguishing between $8$ authors, using five books  to represent each author's style. More specifically, each
book described in Sec.~\ref{ssec.books} was characterized by the set of $15$ features discussed in Sec. \ref{ssec.localtoglobal} ($\langle X
\rangle$, $\langle X \rangle_2$ and $\gamma(X)$ for $X = \{N, B, L, C, I\}$). The authorship assignment was performed using the algorithms
in Sec.~\ref{mlalg} applied to a training dataset independent of the test book using the cross validation methodology (see SI-Sec.~4). This
technique ensures that the training and evaluation sets are different and it is equivalent to assigning the
authorship of one book in an experiment where
$4$ books of $8$ authors were used as a training dataset. The final output of the algorithms is the assignment of a specific author to each book tested, and the efficiency is quantified simply as the fraction of successful assignments.

The results are summarized in Table ~\ref{tab.accuracy} and indicate accuracy rates between $42.5\%$ and $50.0\%$ when all $15$ features were used. These results were statistically significant by a large amount, confirming that these features successfully capture author specific characteristics. To further explore the accuracy of different methods, we considered cases in which only some of the features were included in the algorithms. We tested all~$2^{15}=32,768$ combinations of the~$15$ features and obtained a best result of~$65.0\%$ of correct assignments.

\begin{table}
\centering
\caption{\label{tab.accuracy} Accuracy rate achieved for the three machine learning algorithms using all $15$ features and the best
  combination of these features. The accuracy is estimated based on~$40$ authorship assignments. The
  p-values correspond to the probability of getting by chance a higher or equal accuracy in one (all features) and in $2^{15}=32,768$ (best
  case) trials. The features included in the best cases can be found in SI-Tables S1-S4.}
\begin{tabular}{@{}l||c c c}
\hline & & Algorithms  &  \\
& Decision Tree C4.5 & Nearest neighbor kNN & Naive Bayes  \\
\hline
All $15$ features & $50.0 \;\%\; (p=1 \; 10^{-8})$  & 	$47.5 \; \%  \; (p=6 \; 10^{-8})$ & $42.5 \; \% \; (p=2 \; 10^{-6})$\\
Best case &$62.5 \;\%\; (p=5 \; 10^{-9})$  & 	$65.0 \; \%  \; (p=4 \; 10^{-10})$ & $62.5 \; \% \; (p=5 \; 10^{-9})$\\
\hline
\end{tabular}
\end{table}

\subsection{Relative importance of different features} \label{analisefatorial}

\begin{table}
\centering
\caption{\label{tab.rankey} Ranking of features based on the accuracy rate of the classifiers, where 1 in the table means best, 2 second best and so on. The results for each classifier algorithm (C4.5, kNN and Bayes) are reported using different ranking procedures combined with multiple features (Mann-Whitney U test, columns 1-3, information gain (column 4) and accuracy using only one feature (columns 5-7)).  The last column reports the Pearson correlation between each feature and the vocabulary size $M$ (number of different words) calculated over the $40$ books in our database. The features in the table are ordered according to the decreasing geometric mean of the ranks obtained in the $3$ multiple features analysis (this ordering is the same achieved by considering for each feature the likelihood of reaching by chance a ranking as good as the one in each of the three ranking schemes). The areas under the curve in the multiple features analysis ranged between $56\%$ and $69\%$.}
\begin{tabular}{@{}l|ccc||cccc||c}
\hline
       & \multicolumn{3}{c||} {\bf Multiple features} & \multicolumn{4}{c||} {\bf Single feature} & Correlation \\
   & C4.5 & kNN & Bayes & Info & C4.5 & kNN & Bayes & with $M$\\
\hline
$\langle N \rangle_2$ & 6 & 1 & 1 & 3 & 2 & 5 & 1 & -0.90\\
$\gamma(I)$	          & 2 & 2 & 2 &10 &12 &9 &	10& -0.08\\
$\langle N \rangle$   & 1 &	6 & 3 &2 &1  &2  & 3 & -0.96\\
$\langle L \rangle$  	  & 7 & 4 & 6 &9  &5  &3  & 8 & 0.85\\
$\langle B \rangle$	  & 5 & 8 & 5 &1  &3  &1  & 2 & 0.98\\
\hline
$\langle I \rangle_2$		  &10 & 3 &10 &15 &15 &12 &12 & -0.34\\
$\langle L \rangle_2$  &8  & 7 &8  &8  &5  &7  &5  & 0.85 \\
$\langle C \rangle$		  &12 &11 &4  &6  &5  &5  &5  & -0.87 \\
$\gamma(L)$		  &4  &13 &11 &13 &10 &9 &13 & -0.13\\
$\gamma(B)$		  &3  &14 &14 &11 &8  &9  &9 & -0.07\\
\hline
$\langle B \rangle_2$	  &9  &9  &9  &7  &9  &14 &5 & 0.88\\
$\langle C \rangle_2$	  &11 &10 &7  &5  &4  &3  &4 & -0.87\\
$\langle I \rangle$	     &13 &5  &12 &12 &13 &15 &10 & -0.29\\
$\gamma(N)$	  &15 &12 &13 &4  &10 &8  &13 & 0.81 \\
$\gamma(C)$    &14 &15 &15 &14 &14 &12 &15 & 0.07 \\
\hline
\end{tabular}
\end{table}

In evaluating the importance of the different features on the final results it is essential to identify their mutual dependency. We start from the list of all~$2^{15}=32,768$ combinations of features ordered by decreasing accuracy (as shown in SI-Tables S1-S4). We wish to quantify when feature $y$ appears in the top of this list. To this end, we count the fraction of the $2^{14}$ feature combinations that include $y$ with accuracy higher or equal to a threshold. The final
Estimate is then given by the area-under-the curve of the ROC
plot obtained by varying the threshold~\cite{spackman}. This procedure is equivalent to the Mann-Whitney U test~\cite{meanu}. The motivation for using this method is that it evaluates the  importance of a specific feature by taking into account how it combines with the other features to improve the accuracy of the prediction. The method depends both on the prediction algorithm and on the other features.

The features were ranked based on the method described above. The results for the $3$ prediction algorithms are given in the three first
columns of Table ~\ref{tab.rankey}. The three features appearing as the most prominent are $\langle N \rangle$ (average frequency),
$\gamma (I)$ (skewness of intermittency), and $\langle N \rangle_2$ (average logarithmic frequency).
In order to state the importance of features beyond specific algorithm it is important to quantify in which extent the results obtained for
the three algorithms (first 3 columns) are consistent with each other. To this end we compute the Spearman's rank correlation and obtain the
values 0.29 (p-value $= 0.145$), 0.49 (p-value$ = 0.032$) and 0.67 (p-value$ = 0.003$) for the pairs C4.5/kNN, C4.5/Bayes and
kNN/Bayes, respectively. The p-values are computed under the null hypothesis that the rankings are independent. Altogether, the three
p-values indicate that the three rankings are consistent with each other. This is a strong indication that our analysis goes beyond
algorithm-specific results and indeed captures the influence from the features.

It is interesting to compare the results to evaluations taking into account each feature separately. This can be
done either by considering the accuracy of the prediction using only the specific feature or by comparing the information gained by including
the feature~\cite{infogain}. This last method has the advantage of being independent of the prediction
algorithm. These results are shown in the 4 last columns of Table ~\ref{tab.rankey}. Note that some features appearing as very
important in the multiple features analysis are not informative when taken alone (e.g., the skewness of the intermittency $\gamma(I)$). On
the other hand, features that are well ranked in the single feature analysis do not always appear among the most important features when multiple
features are considered (e.g., the weighted average of the clustering $\langle C \rangle_2$). These observations show the nontrivial
mutual dependency of the features. To further explore this we performed a factorial analysis (see SI-Sec.~5) using the 12 most important features in Table \ref{tab.rankey}, with the most important combinations being summarized in Table \ref{tab.fatorial}.  As expected from Table \ref{tab.rankey}, in fact $\gamma(I)$ appears among the 2 best combinations of features in all three algorithms, which confirms that its effectiveness is correlated with its interdependence with other features.

\begin{table}
\centering
\caption{\label{tab.fatorial} List of the 2 most relevant combinations of features, as revealed by a Factorial Analysis for the C4.5, kNN and Bayes classifier algorithms. As expected from Table \ref{tab.rankey}, $\gamma(I)$ provides good results when used in conjunction with other features, such as $\langle N \rangle$, $\langle N \rangle_2$ $\gamma(B)$ and $\langle I \rangle_2$ }.
\begin{tabular}{@{}l|c|c|c}
\hline
		&	C4.5	&	kNN	&	Bayes \\
\hline
1st Combination	&	$\gamma(L)$, $\langle C \rangle$ and $\langle C \rangle_2$ 				& $\gamma(I)$ and $\gamma(B)$		&	$\langle N \rangle$, $\langle N \rangle_2$ and $\langle B \rangle$ \\
2nd Combination	&  $\gamma(I)$ and $\langle N \rangle$								  				& $\gamma(I)$ and $\langle N \rangle_2$	&  $\langle I \rangle_2$ and $\gamma(I)$ \\
\hline
\end{tabular}
\end{table}

\section{Discussion and conclusions} \label{conclusion}\label{sec.4}

\subsection{Interpretation of the results }

We are now in a position to use the word-specific analysis (Sec.~\ref{sec.2}) and the distribution (Sec.~\ref{ssec.localtoglobal}) of the quantities
$X=\{N,B,L,C,I\}$ to assess the importance of the different features $\langle X \rangle, \langle X \rangle_2$, and $\gamma(X)$ in
Table ~\ref{tab.rankey}:

\begin{itemize}

\item[$N$] frequency. This was the most efficient quantity for recognizing authorship with $\langle N \rangle$ and $\langle N \rangle_2$ among the $3$ most important features.  Noting that~$\langle N \rangle$
is proportional to the inverse number of distinct words~$M$:
\begin{equation}
	\langle N \rangle = \frac{\text{length of book}}{M},
\end{equation}
one infers that the distinguishing feature between the authors captured by $\langle N \rangle$ is the different vocabulary sizes. The
modified average  $\langle N \rangle_2=\langle \log N \rangle$ also captures this aspect, including the proportion of frequent and
infrequent words. On the other   hand, the poor performance of $\gamma(N)$ ($= \alpha$ in $\rho(N)\sim N^{-\alpha}$) is a clear signature of
the universal,  author-independent, character of Zipf's law (at fixed book size~\cite{bernhardsson}).

\item[$B$] betweenness. The average betweenness $\langle B \rangle$ was useful because of its strong correlation with the vocabulary size of the book~$M$ (last column in Table ~\ref{tab.rankey}).
In network terms, this corresponds to a linear relationship between~$\langle B \rangle$ and network size ($M$, number of nodes) and   can be understood by noting that the number of terms in the sum of  definition of $B_i$ in Eq.~(\ref{eq.bi}) is proportional to~$M^2$,   so that $\langle B \rangle$ is expected to scale linearly with~$M$ for a fixed book size. The fact that $\langle B \rangle_2$ and $\gamma(B)$ had a poor   performance indicates that the number of words with high betweenness is not a relevant feature to distinguish between authors.

\item[$L$] shortest path. This was the network quantity with best performance. $\langle L \rangle$ quantifies the   typical distance of words to the central hubs of the network (frequent words). The good performance points to a dependence on the style of the authors. The poorer performance of~$\langle L \rangle_2$ and~$\gamma(L)$ indicates that the style dependency in~$L$ is more prominent in the typical values than, respectively, in the frequent and large~$L$ words.

\item[$C$] clustering. The poor performance of all values related to this quantity suggests that authors have very little freedom in
  choosing   the clustering of words co-occurrence networks. The last position in the ranking of $\gamma(C)$ in Table ~\ref{tab.rankey}
  suggests that the fraction of words used in specific contexts (high $C$) is author independent. The two averages~$\langle C \rangle_2$
  and~$\langle C \rangle$ take similar values (as seen in Fig.~\ref{fig4} and in SI-Sec.~3, recall the restriction $N_i\ge 5$ used  in
  Sec.~\ref{sec.network}). They perform well only when used alone, possibly because of their correlation to vocabulary size~$M$.

\item[$I$] intermittency. Apart from the frequency, intermittency was the most important quantity in Table~\ref{tab.rankey} with the
  skewness of the distribution~$\gamma(I)$ playing a prominent role. In view of the results in Sec.~\ref{sec.intermittency}, $\gamma(I)$ may
  be interpreted as the fraction of all words that are topical or ``keyword like''. The poor performance of $\langle I \rangle$ is not
  surprising since $I$ is normalized by frequency ($I\equiv \sigma_T/\overline{T}$) and therefore $\langle I \rangle \approx 1$ is expected. Indeed, from all $5$ quantities the $I-$features have shown altogether the smallest absolute value of correlation with vocabulary size (Tab.~\ref{tab.rankey}), explaining why even~$\gamma(I)$ has a poor relative performance when used alone. Finally,~$\langle I \rangle_2$ performs better than $\langle I \rangle$ suggesting that frequent words are the more relevant ones.
\end{itemize}

\subsection{Comparison with other prediction methods}\label{ssec.comparison}

Even if the main goal of this paper is to evaluate the importance of different factors, it is also useful to compare the accuracy of our results with other methods of authorship attribution. Uzuner and Katz ~\cite{uzuner} used a database of books similar to ours, produced by $8$ authors. They used five sets of features, including simple statistics and more sophisticated syntactic analysis (Table 3 of Ref.~\cite{uzuner}). Our best results (accuracy of $65\%$) is comparable to their second best case obtained using ``syntactic elements of expression'' ($62\%$), being significantly worse only than their best result, achieved
using function words ($87\%$).
In an extensive review, Grieve ~\cite{Grieve} reported accuracies obtained with a set of $34$ features varying between $33\%$ and $87\%$ for the case of $5$ authors, and between $18\%$ and $80\%$ for $10$ authors (Table 9 in Ref.~\cite{Grieve}). Our best results are above the median of their results achieved by using different features. Their best results again are based on the relative frequency of function words. These results are in accordance with the long tradition started by
Mosteller and Wallace to use the frequency of function words to distinguish between authors~\cite{Mosteller}.

In order to confirm this in our database, we implemented a series of prediction schemes using the frequency $N_i$ of frequent (mostly function) words.  Differently from the approach described in this paper that used average and scaling properties as features, now the frequencies of specific words are used directly as input features of the prediction algorithms. We used only the kNN algorithm because the other algorithms did not provide good results when too many features were included. When the list of $70$ stopwords from Table ~2.5 of Ref.~\cite{Mosteller} was used, we obtained an accuracy of $62.5\%$, i.e., comparable to our best
results. Following Ref.~\cite{Grieve}, we considered two other lists of words: all $1,978$ words that appear in at least one book of
each of the authors, leading to accuracy of $90\%$; and all $209$ words that appear at least once in every book in our database, leading to
an accuracy of $82.5\%$.
We recall that in order to concentrate on analysis that focus on words with pronounced semantic content instead of function words, we have deliberately excluded a list of stopwords that comprised $80\%$ of the cases listed in Ref.~\cite{Mosteller}. Therefore, our best combination of features compares well to other methods which demand more sophisticated syntactic analysis of the text. Measurements of complex network and intermittency are indeed able to capture many of the author-dependent characteristics.

In order to illustrate how measures analyzed in this paper can be complementary to traditional methods we have performed a very simple
experiment using as features the frequency and intermittency of the set of words composed by the five most frequent words in each book.  The
accuracy in classifying the books only by
the frequency was $72.5~\%$ and only by intermittency was $37.5~\%$. Although this last accuracy rate is not impressive, it is statistically
significant ($p = 2.2~10^{-4}$) and shows that the intermittency values of specific words across distinct authors is different. The accuracy
is increased to $80~\%$ when both features were included. It remains to be shown in future works how our results can improve state of the art
methods of authorship attribution.


\subsection{Summary of conclusions}

We have shown that the style of
different authors leave fingerprints in very general statistical measures of texts based on the network of co-occurrence of words and on
intermittency or burstiness of words. The statistically significant scores obtained in authorship attribution unequivocally show that the
style dependence of these features can be used in practice. Regarding the prominence of the different features, we note that both the
results and ranking of features may depend on the database, selected features and attribution algorithms. Accordingly, as emphasized in
Ref.~\cite{Grieve}, different algorithms and features have to be tested in a given corpus before any real application of authorship
attribution. However, the robustness of our results using three radically different attribution algorithms strongly suggests that the
different features have importance that go beyond
specific algorithms. Two features should be highlighted: (i) the skewness of the distribution of intermittent words
$\gamma(I)$, which is based on the long-scale distribution of words and detects the extent into which topical words (keywords,
large $I\equiv \sigma_T/\overline{T}$) were used in the book; and (ii) the mean shortest path of the word co-occurrence network
$\langle L \rangle$, which is based on the short-range connectivity of words and detects the typical distance of words to all other words. The different natures of these two quantities suggest a complementary role for capturing both short- and long-scale properties of the text, as well as typical and exceptional words.

Our focus in this paper was on the evaluation of the different features, rather than on maximizing the efficiency of the authorship attribution
algorithms. This is apparent when comparing the best accuracy rates we achieved using our approach ($62.5\%$) and using previous
proposals ($90.0\%$), as discussed in Sec.~\ref{ssec.comparison}. A further limitation of approaches based on intermittency and networks is that they require large pieces of text.
While the root of the success of previous methods rely on the observation that
function words are a powerful tool to detect the style of authors~\cite{Mosteller,Grieve,uzuner},
in the complex network and intermittency approaches used in this paper the focus is on the content words. In this sense the results we achieve can be thought as being complementary to the analysis using function words. More specifically, our results suggest that using $\gamma(I)$ and $\langle L \rangle$ can improve authorship recognition techniques when used in combination with the many different features currently employed~\cite{Grieve}. Finally, the successful application of these measurements to characterize the style of authors suggests that the quantities discussed here can be further explored in other linguistic tasks, an approach that has been limited to a few works (see e.g.~\cite{iss,dmitry}).





\end{document}